\documentstyle[12pt]{article}%
\newlength\Cscr\newlength\Csave
\newlength\Ctenthex\newlength\CFxsize
\newlength\CFxsizeps\newlength\CFsizemakebox\newlength\CFleftcrop
\newlength\CFrightcrop\newlength\CZtbldist\newlength\CZfigdist
\newlength
\CGDnum\newlength\CGDtext%
\newcounter{CBauthornu%
m}\newcounter{Ceqindent}\newcounter{CBtnc}
\newcounter{CBtntc}\newcounter{CEht}%
\newcounter{Cbscurr}\newcounter{CbsA}\newcounter{CbsB}\newcounter
{CbsC}\newcounter{CbsD}\hoffset\Cscr\voffset
\Cscr\protect
\begin{document}\setcounter{figure}{1}\setcounter{t%
able}{0}\renewcommand\theequation{\arabic{equation}}\renewcommand
\thetable{\arabic{table}}\renewcommand\thefigure{\arabic{figure}%
}\renewcommand\thesection{\arabic{section}}\renewcommand
\thesubsection{\arabic{section}.\arabic{subsection}}\renewcommand
\thesubsubsection{\arabic{section}.\arabic{subsection}.\arabic{s%
ubsubsection}}\setcounter{CEht}{10}\setcounter{CbsA}{1}%
\setcounter{CbsB}{1}\setcounter{CbsC}{1}\setcounter{CbsD}{1}{%
\centering{\protect\mbox{}}\\*[\baselineskip]{\large\bf A new pr%
oposal for the fermion doubling problem}\\*}\addtocounter{CBtntc%
}{1}{\centering{\protect\mbox{}}\\John P.~Costella\vspace{1ex}%
\\*}{\centering{\small\mbox{}\protect\/{\protect\em Faculty of M%
athematics, Mentone Grammar, 63 Venice Street, Mentone, Victoria 
3194, Australia\protect\/}}\\}{\centering{\small\mbox{}\protect
\/{\protect\em jpcostella@hotmail.com;\hspace{1ex} jpc@mentonegs%
.vic.edu.au;\hspace{1ex} jpc@physics.unimelb.edu.au\protect\/}}%
\\}{\centering{\protect\mbox{}}\\(19\ July 2002\vspace{1ex})\\}%
\par\vspace\baselineskip\setlength{\Csave}{\parskip}{\centering
\small\bf Abstract\\}\setlength{\parskip}{-\baselineskip}\begin{%
quote}\setlength{\parskip}{0.5\baselineskip}\small\noindent In t%
his paper I propose the use of a lattice derivative operator tha%
t is equivalent to the ideal SLAC derivative operator in all lat%
tice calculations, but without the prohibitively expensive compu%
tational cost of the latter. A pedagogical motivation and deriva%
tion of the closed form of the SLAC derivative in position space 
is presented, and the proposed method for its cost-effective imp%
lementation is presented in detail. \end{quote}\setlength{%
\parskip}{\Csave}\par\refstepcounter{section}\vspace{1.5%
\baselineskip}\par{\centering\bf\thesection. Pedagogical overvie%
w of the fundamental reasons for fermion doubling\\*[0.5%
\baselineskip]}\protect\indent\label{sect:Doubling}Perhaps the m%
ost important ingredient in any good lattice calculation is the 
fundamental building-in of as many of the symmetries of the cont%
inuum formalism being modelled as possible, so that the results 
respect the given symmetries identically, rather than only in th%
e continuum limit. However, the conflicting requirements of thes%
e symmetries may, in turn, introduce subtle artefacts into the f%
ormalism that destroy its physical correspondence; and these art%
efacts can in some cases be difficult to exorcise.\par The fermi%
on doubling problem is a particularly notorious example of such 
an artefact. It arises from the apparently simple requirement th%
at we construct a spatial derivative operator on the lattice---e%
ven in one dimension. In the first instance, one might think tha%
t the first-principles definition of the derivative taught to sc%
hool children, \setcounter{Ceqindent}{0}\protect\begin{eqnarray}%
\protect\left.\protect\begin{array}{rcl}\protect\displaystyle
\hspace{-1.3ex}&\protect\displaystyle\mbox{$\protect\displaystyle
\protect\frac{df}{dx}$}\equiv\lim_{a\rightarrow0}\mbox{$\protect
\displaystyle\protect\frac{f(x+a)-f(x)}{a}$},\setlength{\Cscr}{%
\value{CEht}\Ctenthex}\addtolength{\Cscr}{-1.0ex}\protect
\raisebox{0ex}[\value{CEht}\Ctenthex][\Cscr]{}\protect\end{array%
}\protect\right.\protect\label{eq:Doubling-FirstPrinciples}%
\protect\end{eqnarray}\setcounter{CEht}{10}might be an ideal can%
didate for the lattice derivative operator: we need simply take 
$a$ to be the lattice spacing, and the limit will automatically 
be taken when we extrapolate to the continuum limit of the calcu%
lations. In fact, in a huge number of computational applications 
in engineering, such a definition is perfectly acceptable, and i%
s used every day without complication.\par The problem arises if 
we wish to perform any calculation in quantum physics. Simple no%
nrelativistic quantum mechanics is enough to highlight the diffi%
culties. There, we wish to construct a free-particle momentum op%
erator \mbox{$\protect\displaystyle p\equiv-i^{\:\!}d/dx$}. (Thr%
oughout this paper, I use units in which \mbox{$\protect
\displaystyle\hbar=1$}.) However, since we want the momentum to 
be an observable, its operator representation must (by the princ%
iples of quantum mechanics) be Hermitian. This, in turn, implies 
that the operator \mbox{$\protect\displaystyle d/dx$} must be 
\mbox{}\protect\/{\protect\em anti-Hermitian\protect\/}. What do%
es this mean in the context of a one-dimensional lattice calcula%
tion? Think of the nonrelativistic wavefunction $\psi(x)$ as a c%
olumn vector, each element of which is a complex number, represe%
nting the value of the wavefunction at the corresponding lattice 
site. The conjugate wavefunction \mbox{$\protect\displaystyle\psi
^{\ast\:\!\!}(x)$} can be represented as the Hermitian conjugate 
of this column vector (which we should probably denote \mbox{$%
\protect\displaystyle\psi^{\dagger\:\!\!}(x)$} instead). Operato%
rs are sandwiched between $\psi^\dagger$ and $\psi$, so must be 
matrices; each matrix element determines how $\psi$ at some $x$ 
is coupled to \mbox{$\protect\displaystyle\psi^{\ast\!}$} at som%
e other $x'$\mbox{$\!$}.\par Anti-Hermiticity of the derivative 
operator then simply requires that its matrix representation be 
anti-Hermitian. The first-principles prescription (\protect\ref{%
eq:Doubling-FirstPrinciples}), however, corresponds to an operat%
or of the form \setcounter{Ceqindent}{0}\protect\begin{eqnarray}%
\hspace{-1.3ex}&\displaystyle\mbox{$\protect\displaystyle\protect
\frac{1}{a}$}\!\protect\left(\begin{array}{ccccc}\!\!-1\!&1&0&0&%
\cdots\\[0.5ex]0&\!\!-1\!&1&0&\cdots\\[0.5ex]0&0&\!\!-1\!&1&%
\cdots\\\vdots&\vdots&\vdots&\!\ddots\!&\ddots\end{array}\protect
\right)\!,\protect\nonumber\setlength{\Cscr}{\value{CEht}%
\Ctenthex}\addtolength{\Cscr}{-1.0ex}\protect\raisebox{0ex}[%
\value{CEht}\Ctenthex][\Cscr]{}\protect\end{eqnarray}\setcounter
{CEht}{10}which is clearly not antisymmetrical, as is required i%
f a real matrix is to be anti-Hermitian.\par A decision must the%
refore be made. Either one must give up the simple derivative op%
erator (\protect\ref{eq:Doubling-FirstPrinciples}), or else one 
must give up the Hermiticity of the momentum operator. The latte%
r course of action would, of course, destroy the unitarity of th%
e formalism; and, even though unitarity would (presumably) be re%
trieved in the continuum limit, the high importance afforded to 
this fundamental ``symmetry'' of quantum physics generally overr%
ides any thought of doing so.\par Thus, one generally chooses th%
e former alternative, namely, the rejection of the first-princip%
les operator (\protect\ref{eq:Doubling-FirstPrinciples}). Howeve%
r, before we flush it away, we should first be aware of what we 
are discarding. Consider the discrete Fourier transform (namely, 
the momentum-space representation) of $-i$ times the operation r%
epresented by (\protect\ref{eq:Doubling-FirstPrinciples}): 
\setcounter{Ceqindent}{0}\protect\begin{eqnarray}\protect\left.%
\protect\begin{array}{rcl}\protect\displaystyle\hspace{-1.3ex}&%
\protect\displaystyle p_{\mbox{\scriptsize$\:\!$first-principles%
}}=\mbox{$\protect\displaystyle\protect\frac{e^{ipa}-1}{ia}$}%
\equiv e^{ipa/2}\,\mbox{$\protect\displaystyle\protect\frac{e^{i%
pa/2}-e^{-ipa/2}}{ia}$}\equiv\mbox{$\protect\displaystyle\protect
\frac{2}{a}$}\sin\!\protect\left(\mbox{$\protect\displaystyle
\protect\frac{pa}{2}$}\protect\right)e^{ipa/2}.\setlength{\Cscr}%
{\value{CEht}\Ctenthex}\addtolength{\Cscr}{-1.0ex}\protect
\raisebox{0ex}[\value{CEht}\Ctenthex][\Cscr]{}\protect\end{array%
}\protect\right.\protect\label{eq:Doubling-PFirstPrinciples}%
\protect\end{eqnarray}\setcounter{CEht}{10}The final factor of $%
e^{ipa/2}$ is simply a phase factor, which encapsulates the non-%
Hermiticity of the operator---namely, the reason we wish to reje%
ct it. Let us ignore it for the moment---say, by defining 
\setcounter{Ceqindent}{0}\protect\begin{eqnarray}\hspace{-1.3ex}%
&\displaystyle\tilde p_{\mbox{\scriptsize$\:\!$first-principles}%
}\equiv\mbox{$\protect\displaystyle\protect\frac{2}{a}$}\sin\!%
\protect\left(\mbox{$\protect\displaystyle\protect\frac{pa}{2}$}%
\protect\right)\!.\protect\nonumber\setlength{\Cscr}{\value{CEht%
}\Ctenthex}\addtolength{\Cscr}{-1.0ex}\protect\raisebox{0ex}[%
\value{CEht}\Ctenthex][\Cscr]{}\protect\end{eqnarray}\setcounter
{CEht}{10}This latter operator is perfectly well-defined. For sm%
all $p$, we find that \mbox{$\protect\displaystyle\tilde p_{\mbox
{\scriptsize$\:\!$first-principles}}\approx p$}, as would be exp%
ected for any reasonable momentum operator. At the Brillouin zon%
e boundary, namely, at \mbox{$\protect\displaystyle p=\pm\pi/a$}%
, we find \mbox{$\protect\displaystyle\tilde p_{\mbox{\scriptsize
$\:\!$first-principles}}=\pm2/a$}; and $2/a$ is a large, nonzero 
value. The shape of the function is not quite right: instead of 
being a linear (namely, just $p$), it ``bends over'' more and mo%
re as one approaches the Brillouin zone boundary, and is station%
ary at the boundary. But it is nevertheless monotonic in $p$, an%
d has no zeroes except at \mbox{$\protect\displaystyle p=0$}.\par
How does the first-principles momentum operator manage to have a 
discontinuity across the Brillouin zone boundary? It doesn't, of 
course: we have ignored the extra phase factor $e^{ipa/2}$, whic%
h has the effect of ``twisting'' the function away from the real 
axis: at the Brillouin zone boundary, it has twisted \mbox{$%
\protect\displaystyle p_{\mbox{\scriptsize$\:\!$first-principles%
}}$} around to be purely imaginary, with the $\pm90\mbox{$^\circ
$}$ ``twist'' for \mbox{$\protect\displaystyle p=\pm\pi/a$} brin%
ging the two ends of the function together.\par In any case, we 
are rejecting this first-principles operator, on the grounds tha%
t we wish to preserve unitarity. What next? Obviously, we want a 
definition of the derivative that is more symmetrical about the 
position in question than (\protect\ref{eq:Doubling-FirstPrincip%
les}). The obvious thing to try is \setcounter{Ceqindent}{0}%
\protect\begin{eqnarray}\protect\left.\protect\begin{array}{rcl}%
\protect\displaystyle\hspace{-1.3ex}&\protect\displaystyle\mbox{%
$\protect\displaystyle\protect\frac{df}{dx}$}\equiv\lim_{a%
\rightarrow0}\mbox{$\protect\displaystyle\protect\frac{f(x+a)-f(%
x-a)}{2a}$},\setlength{\Cscr}{\value{CEht}\Ctenthex}\addtolength
{\Cscr}{-1.0ex}\protect\raisebox{0ex}[\value{CEht}\Ctenthex][%
\Cscr]{}\protect\end{array}\protect\right.\protect\label{eq:Doub%
ling-Naive}\protect\end{eqnarray}\setcounter{CEht}{10}again with 
$a$ being taken to be the lattice spacing. The matrix correspond%
ing to this operator now has the desired antisymmetry: 
\setcounter{Ceqindent}{0}\protect\begin{eqnarray}\hspace{-1.3ex}%
&\displaystyle\mbox{$\protect\displaystyle\protect\frac{1}{2a}$}%
\!\protect\left(\begin{array}{cccccc}0&1&0&0&0&\cdots\\[0.5ex]\!%
\!-1\!&0&1&0&0&\cdots\\[0.5ex]0&\!\!-1\!&0&1&0&\cdots\\[0.5ex]0&%
0&\!\!-1\!&0&1&\cdots\\\vdots&\vdots&\vdots&\!\ddots\!&\!\ddots
\!&\ddots\end{array}\protect\right)\!.\protect\nonumber\setlength
{\Cscr}{\value{CEht}\Ctenthex}\addtolength{\Cscr}{-1.0ex}\protect
\raisebox{0ex}[\value{CEht}\Ctenthex][\Cscr]{}\protect\end{eqnar%
ray}\setcounter{CEht}{10}(There is a $-1$ element in the top-rig%
ht corner, and $+1$ in the bottom-left, if we want to preserve t%
ranslational invariance by imposing periodic boundary conditions%
.) It is remarkable---even if usually taken for granted---that t%
he changes in ``epsilontics'' represented by the change from (%
\protect\ref{eq:Doubling-FirstPrinciples}) to (\protect\ref{eq:D%
oubling-Naive}) can make the difference between violating and ob%
serving the unitarity of the underlying formalism.\par It is the 
definition (\protect\ref{eq:Doubling-Naive}) that formed the bas%
is of most early lattice calculations. However, this operator is 
still very na\-\"\i ve\ in its structure. (It is, for example, a%
lso taught to school students.) Moreover, by trying to preserve 
unitarity, in a na\-\"\i ve\ way, we have immediately introduced 
spurious properties into the derivative operator that are comple%
tely responsible for the fermion doubling problem. It is straigh%
tforward to see why this is the case. In the definition (\protect
\ref{eq:Doubling-Naive}), we are no longer taking a finite diffe%
rence between adjacent lattice sites, but we are rather \mbox{}%
\protect\/{\protect\em skipping\protect\/} a site, and comparing 
the function at one site to the function two sites away from it. 
Now, recall that the highest-frequency normalised wave that can 
be represented on the lattice (at the boundary of the first Bril%
louin zone) simply oscillates \mbox{$\protect\displaystyle+1,-1,%
+1,-1,\ldots$} as we move along the lattice sites. The first-pri%
nciples operator (\protect\ref{eq:Doubling-FirstPrinciples}) com%
pares adjacent sites, and finds a large (indeed, maximal) change 
of $\pm2$. The na\-\"\i ve\ operator (\protect\ref{eq:Doubling-N%
aive}), on the other hand, skips every other site, and compares 
$+1$ with $+1$, and $-1$ with $-1$, and concludes that the deriv%
ative of the wave is zero everywhere!\par This vanishing of the 
na\-\"\i ve\ derivative operator at the Brillouin zone boundary 
would not, in itself, be problematical, if it were contained to 
only this one frequency, because the anomalous properties of one 
single mode out of the total number of momentum modes (equal to 
the number of lattice sites) would lead to negligible effects as 
the number of lattice sites is increased. Rather, the real probl%
em arises in the region \mbox{}\protect\/{\protect\em near%
\protect\/} the Brillouin zone boundary. If we take the discrete 
 Fourier transform of the operator (\protect\ref{eq:Doubling-Nai%
ve}), we find \setcounter{Ceqindent}{0}\protect\begin{eqnarray}%
\protect\left.\protect\begin{array}{rcl}\protect\displaystyle
\hspace{-1.3ex}&\protect\displaystyle p_{\mbox{\scriptsize na\-%
\"\i ve}}=\mbox{$\protect\displaystyle\protect\frac{e^{ipa}-e^{-%
ipa}}{2ia}$}\equiv\mbox{$\protect\displaystyle\protect\frac{1}{a%
}$}\sin pa.\setlength{\Cscr}{\value{CEht}\Ctenthex}\addtolength{%
\Cscr}{-1.0ex}\protect\raisebox{0ex}[\value{CEht}\Ctenthex][\Cscr
]{}\protect\end{array}\protect\right.\protect\label{eq:Doubling-%
PNaive}\protect\end{eqnarray}\setcounter{CEht}{10}As promised, t%
he unitarity-breaking phase factor of (\protect\ref{eq:Doubling-%
PFirstPrinciples}) is absent; and for small $p$, we find \mbox{$%
\protect\displaystyle p_{\mbox{\scriptsize na\-\"\i ve}}\approx
p$}, as desired. However, by skipping a site in position space, 
we have halved the period of the function in momentum space, so 
that instead of approaching a large value at the Brillouin zone 
boundary, $p_{\mbox{\scriptsize na\-\"\i ve}}$ smoothly approach%
es zero! Indeed, the positive-$p$ part of the first Brillouin zo%
ne has effectively been divided into two mirror halves: from 
\mbox{$\protect\displaystyle p=0$} to \mbox{$\protect
\displaystyle p=\pi/2a$}, in which $p_{\mbox{\scriptsize na\-\"\i
ve}}$ increases monotonically, and represents a reasonably faith%
ful representation of the momentum operator; and then from \mbox
{$\protect\displaystyle p=\pi/2a$} to \mbox{$\protect
\displaystyle p=\pi/a$}, in which $p_{\mbox{\scriptsize na\-\"\i
ve}}$ actually \mbox{}\protect\/{\protect\em decreases\protect\/%
} for increasing $p$. (The same description can be made for the 
negative-$p$ part of the first Brillouin zone.)\par It is these 
``mirror states'' that are fundamentally responsible for the fer%
mion doubling problem using the na\-\"\i ve\ derivative operator 
(\protect\ref{eq:Doubling-Naive}). The problem only arises for f%
ermions, because the Dirac operator contains the first-derivativ%
e, whereas for bosons we only require the \mbox{}\protect\/{%
\protect\em second\protect\/}-derivative, which can be reasonabl%
y approximated by Hermitian operator \setcounter{Ceqindent}{0}%
\protect\begin{eqnarray}\protect\left.\protect\begin{array}{rcl}%
\protect\displaystyle\hspace{-1.3ex}&\protect\displaystyle\mbox{%
$\protect\displaystyle\protect\frac{d^{\:\!2}\!f}{dx^2}$}\equiv
\lim_{a\rightarrow0}\mbox{$\protect\displaystyle\protect\frac{f(%
x+a)-2f(x)+f(x-a)}{a^2}$},\setlength{\Cscr}{\value{CEht}\Ctenthex
}\addtolength{\Cscr}{-1.0ex}\protect\raisebox{0ex}[\value{CEht}%
\Ctenthex][\Cscr]{}\protect\end{array}\protect\right.\protect
\label{eq:Doubling-SecondDeriv}\protect\end{eqnarray}\setcounter
{CEht}{10}which does not skip any lattice sites, and hence exhib%
its neither a zero at the Brillouin zone boundary nor the ``mirr%
or states'' phenomenon.\par Thus, by saving unitarity, we have i%
ntroduced spurious properties into the momentum operator, effect%
ively leading to a doubling of the fermion species. It used to b%
e widely believed that, by the Nielson--Ninomiya ``no-go'' theor%
em, such fermion doubling was essentially unavoidable in any lat%
tice formalism of interest to high-energy physics. This belief i%
s, however, founded on a misconception, as will be further discu%
ssed in \mbox{Sec.~$\:\!\!$}\protect\ref{sect:Local}.\par
\refstepcounter{section}\vspace{1.5\baselineskip}\par{\centering
\bf\thesection. The ideal (SLAC) derivative operator\\*[0.5%
\baselineskip]}\protect\indent\label{sect:Ideal}If the na\-\"\i
ve\ derivative operator, (\protect\ref{eq:Doubling-Naive}), suff%
ers pathologically from the doubling problem, then how are we to 
define a derivative operator that still maintains unitarity?\par
Clearly, the ideal situation would be if the discrete Fourier tr%
ansform of the derivative operator were to be simply \setcounter
{Ceqindent}{0}\protect\begin{eqnarray}\protect\left.\protect
\begin{array}{rcl}\protect\displaystyle\hspace{-1.3ex}&\protect
\displaystyle p_{\mbox{\scriptsize$\:\!$ideal}}=p\setlength{\Cscr
}{\value{CEht}\Ctenthex}\addtolength{\Cscr}{-1.0ex}\protect
\raisebox{0ex}[\value{CEht}\Ctenthex][\Cscr]{}\protect\end{array%
}\protect\right.\protect\label{eq:Ideal-PIdeal}\protect\end{eqna%
rray}\setcounter{CEht}{10}(the ``SLAC'' prescription of Drell, W%
einstein and Yankielowicz.) Being real, such an operator is clea%
rly Hermitian, and so unitarity would be preserved. Without spur%
ious zeroes and mirror states in the first Brillouin zone, there 
is no doubling problem. Moreover, the shape of the Fourier trans%
form, namely, perfectly linear, means that the errors involved i%
n performing the derivative on a discrete lattice are minimised; 
such errors are manifested in the ``bending'' of the Fourier tra%
nsform of the derivative error away from linearity.\par There ar%
e several immediate objections to a derivative operator satisfyi%
ng (\protect\ref{eq:Ideal-PIdeal}). The first is that, being a d%
iscontinuous function in momentum space at the Brillouin zone bo%
undary, such an operator much necessarily be ``nonlocal'' in pos%
ition space. This not only causes us concern from a conceptual p%
oint of view---we generally wish to study manifestly local gauge 
theories---but also implies a huge explosion in computational co%
st, because instead of using just two lattice sites to perform a 
derivative operation, we would need to make use of every single 
one of them.\par The conceptual objections to ``nonlocality'' wi%
ll be addressed in \mbox{Sec.~$\:\!\!$}\protect\ref{sect:Local}, 
and the avoidance of the explosion of computational cost will be 
addressed in \mbox{Sec.~$\:\!\!$}\protect\ref{sect:Proposed}. Le%
t us, therefore, put these concerns to one side, for the moment.%
\par A more subtle objection is that it is not immediately obvio%
us what the discrete Fourier transform of (\protect\ref{eq:Ideal%
-PIdeal}) actually is. This does not present any \mbox{}\protect
\/{\protect\em practical\protect\/} barrier, of course, because 
for any particular lattice size, it is a simple enough computati%
onal task to perform the discrete Fourier transform of (\protect
\ref{eq:Ideal-PIdeal}), numerically. However, such a way of proc%
eeding would be conceptually unsatisfactory---to me, at any rate%
---because it would leave quite mysterious the question of what 
the operator corresponding to (\protect\ref{eq:Ideal-PIdeal}) is 
actually doing in position space.\par Fortunately, there is a re%
latively simple trick that allows us to obtain the discrete Four%
ier transform of (\protect\ref{eq:Ideal-PIdeal}) analytically. T%
o explain it in a way that is easy to understand, I must first d%
igress into an analogous physical problem: the case of an engine%
er wanting to sample a signal---for example, in order to digitis%
e an audio track and record it onto a CD. The engineer's ``posit%
ion space'' is actually time, whereas for the lattice we are thi%
nking of spatial dimensions, but the mathematics and the concept%
s are the same.\par Now, the engineer does \mbox{}\protect\/{%
\protect\em not\protect\/} simply pass his audio signal into an 
analogue-to-digital converter, sampling at a suitably high rate. 
If he did, he would, in general, find that there are frequency c%
omponents in the audio signal that are outside the first Brillou%
in zone. (He calls it by a different name, but we won't worry ab%
out that.) When sampled, these higher-frequency components would 
be ``folded back'' into the first Brillouin zone (he calls it ``%
aliasing''); and when the signal is reconstituted in the CD play%
er, they would cause audible interference or degradation (depend%
ing on whether they are coherent sounds or just simply noise).%
\par To avoid this phenomenon, the engineer first passes the aud%
io signal through a ``low-pass filter''\mbox{$\!$}. Ideally, suc%
h a filter would allow through all frequencies in the first Bril%
louin zone, and block all frequencies outside it. In practice, t%
he ideal filter can only be approximated; but we do not care abo%
ut the building of circuits, so we can assume that we have in ou%
r possession the mathematically ideal low-pass filter.\par The e%
ngineer's analysis of this filtering process is as follows. Allo%
wing through the frequencies in the first Brillouin zone, and bl%
ocking those outside it, is equivalent to multiplying the moment%
um spectrum by a filter function that is equal to 1 inside the B%
rillouin zone, and 0 outside it. Multiplying in momentum space i%
s equivalent to convolving in position space, so it is of great 
interest to the engineer to determine the Fourier transform of t%
his low-pass filter. This is easy to do: \setcounter{Ceqindent}{%
0}\protect\begin{eqnarray}\hspace{-1.3ex}&\displaystyle\mbox{$%
\protect\displaystyle\protect\frac{1}{2\pi}$}\hspace{-0.5mm}{%
\protect\mbox{}}\hspace{-0.1mm}\protect\int_{-\pi/a}^{\pi/a}{%
\protect\mbox{}}\hspace{-0.5mm}\hspace{-0.6mm}dp\,e^{ipx}=\mbox{%
$\protect\displaystyle\protect\frac{e^{i\pi x/a}-e^{-i\pi x/a}}{%
2\pi ix}$}\equiv\mbox{$\protect\displaystyle\protect\frac{\sin(%
\pi x/a)}{\pi x}$}.\protect\nonumber\setlength{\Cscr}{\value{CEh%
t}\Ctenthex}\addtolength{\Cscr}{-1.0ex}\protect\raisebox{0ex}[%
\value{CEht}\Ctenthex][\Cscr]{}\protect\end{eqnarray}\setcounter
{CEht}{10}This function plays such a fundamental r\^ole\ in the 
engineer's work that it is given a special name: \setcounter{Ceq%
indent}{0}\protect\begin{eqnarray}\protect\left.\protect\begin{a%
rray}{rcl}\protect\displaystyle\hspace{-1.3ex}&\protect
\displaystyle\mbox{sinc}(x)\equiv\mbox{$\protect\displaystyle
\protect\frac{\sin\pi x}{\pi x}$},\setlength{\Cscr}{\value{CEht}%
\Ctenthex}\addtolength{\Cscr}{-1.0ex}\protect\raisebox{0ex}[%
\value{CEht}\Ctenthex][\Cscr]{}\protect\end{array}\protect\right
.\protect\label{eq:Ideal-Sinc}\protect\end{eqnarray}\setcounter{%
CEht}{10}in terms of which the Fourier transform of the low-pass 
filter can be written \setcounter{Ceqindent}{0}\protect\begin{eq%
narray}\protect\left.\protect\begin{array}{rcl}\protect
\displaystyle\hspace{-1.3ex}&\protect\displaystyle\mbox{$\protect
\displaystyle\protect\frac{1}{a}$}^{\,}\mbox{sinc}\!\protect\left
(\mbox{$\protect\displaystyle\protect\frac{x}{a}$}\protect\right
)\mbox{$\!$}.\setlength{\Cscr}{\value{CEht}\Ctenthex}\addtolength
{\Cscr}{-1.0ex}\protect\raisebox{0ex}[\value{CEht}\Ctenthex][%
\Cscr]{}\protect\end{array}\protect\right.\protect\label{eq:Idea%
l-LowPassX}\protect\end{eqnarray}\setcounter{CEht}{10}The sinc f%
unction has a central peak at \mbox{$\protect\displaystyle x=0$}%
, and oscillates away on each side with an envelope that drops o%
ff like $1/x$. Its fundamental use to the engineer can be seen w%
hen we only consider values of $x$ that are actually on the latt%
ice sites that we wish to use, namely, \mbox{$\protect
\displaystyle x_{n\!}=na$} ($n$ being an integer); in other word%
s, we wish to take the sinc of integer values. For $\mbox{sinc}(%
0)$ we note that \mbox{$\protect\displaystyle\sin\pi x\rightarrow
\pi x$} for small $x$, and hence \mbox{$\protect\displaystyle
\mbox{sinc}(0)=1$}. On the other hand, for all nonzero integral 
values $n$, we find that \mbox{$\protect\displaystyle\mbox{sinc}%
(n)=0$}, because \mbox{$\protect\displaystyle\sin nx=0$} and the 
denominator is nonzero.\par{}From\ (\protect\ref{eq:Ideal-LowPas%
sX}) it can thus be seen that, in position space, the low-pass f%
ilter times the cell width (namely, $a$) is a representation of 
the Dirac delta function on the lattice, with the additional pro%
perty that, when extended to all continuum values of $x$ between 
the lattice sites, it possesses no frequency components outside 
the first Brillouin zone. Thus, when the engineer wishes to reco%
nstitute the analogue sound track from the digitally sampled val%
ues recorded on the CD, he convolves the stream of sampled value%
s with the sinc function---in other words, he places a copy of t%
he sinc function, with a weight given by the amplitude of the so%
und signal at that sample, centred\ on each sample site in quest%
ion; the reconstructed signal is the sum of all of these weighte%
d sinc pulses. The resulting ``smoothly'' reconstructed signal i%
s then guaranteed to have no frequency components outside the fi%
rst Brillouin zone; and, indeed, if it were to be re-sampled at 
the same rate, the fact that the sinc function is zero for all l%
attice sites other than the central one means that there is no 
``crosstalk'': the same sampled values would be obtained. For in%
formation contained within the first Brillouin zone, the entire 
process is completely lossless, without distortion, and reversib%
le.\par So what is the relevance of this interesting digression 
into the world of engineering? It is this: When putting a physic%
al formalism onto a lattice, it is much better (conceptually, at 
least) to break the process into three parts. Firstly, constrain 
the formalism to the first Brillouin zone, using the equivalent 
of a low-pass filter. Secondly, consider what operators will be 
required, and \mbox{}\protect\/{\protect\em construct them expli%
citly\protect\/}, from this low-pass (but still continuum) forma%
lism. Only after this is completed should the third and final st%
ep be taken: the ``sampling'' of the resulting formalism onto th%
e lattice.\par Clearly, this is a general prescription, for the 
placing of any physical formalism whatsoever onto a lattice. The 
particular case of interest to us here is the construction of th%
e spatial derivative operator. So how should we construct it, ac%
cording to this advice?\par First, we need to understand some th%
ings about the derivative operator in the original continuum, un%
filtered formalism. We may consider the process of differentiati%
on to be equivalent to be the effect of convolving the negative 
of the derivative of the Dirac delta function, \mbox{$\protect
\displaystyle-\delta^{\prime\:\!\!}(x)$}, with the function we w%
ish to differentiate, because \setcounter{Ceqindent}{0}\protect
\begin{eqnarray}\hspace{-1.3ex}&\displaystyle-\delta^{\prime\:\!%
\!}\star f(x)\equiv-\hspace{-0.5mm}{\protect\mbox{}}\hspace{-0.1%
mm}\protect\int_{-\infty}^{\infty}{\protect\mbox{}}\hspace{-0.5m%
m}\hspace{-0.6mm}dw\,\delta^{\prime\:\!\!}(x-w)f(w)=+\hspace{-0.%
5mm}{\protect\mbox{}}\hspace{-0.1mm}\protect\int_{-\infty}^{%
\infty}{\protect\mbox{}}\hspace{-0.5mm}\hspace{-0.6mm}dw\,\delta
(x-w)f^{\prime\:\!\!}(w)\equiv f^{\prime\:\!\!}(x),\protect
\nonumber\setlength{\Cscr}{\value{CEht}\Ctenthex}\addtolength{%
\Cscr}{-1.0ex}\protect\raisebox{0ex}[\value{CEht}\Ctenthex][\Cscr
]{}\protect\end{eqnarray}\setcounter{CEht}{10}where in the middl%
e step we have integrated by parts and noted that the surface te%
rm vanishes on account of the delta function. (We are assuming, 
as usual, that the functions $f(x)$ that we wish to act on are s%
ufficiently analytical for their product with the Dirac delta fu%
nction or its derivative to be unambiguous and well-defined.) Th%
us, the derivative operator is, effectively, the negative of the 
derivative of the Dirac delta function. Now, the Fourier transfo%
rm of the Dirac delta function is just unity, and it is an eleme%
ntary property of the Fourier transform that taking the derivati%
ve in position space is simply equivalent to multiplying by $-ip%
$ in momentum space; thus, the Fourier transform of the derivati%
ve operator is simply $ip$, or in other words the Fourier transf%
orm of \mbox{$\protect\displaystyle-i^{\:\!}d/dx$} is just $p$, 
which is exactly what we know from nonrelativistic quantum mecha%
nics.\par Let us now apply the low-pass filter, to allow through 
only that part of the formalism contained within the first Brill%
ouin zone. As noted above, the Fourier transform of the Dirac de%
lta function is unity, but we now filter this so that it is only 
unity within the first Brillouin zone, and zero outside; the Fou%
rier transform of this function is, of course, the sinc function 
described by (\protect\ref{eq:Ideal-LowPassX}).\par Our above an%
alysis then tells us that the derivative operator in the low-pas%
s formalism should simply be taken to be the negative of the der%
ivative of the delta function in the low-pass formalism, namely, 
the negative of the derivative of (\protect\ref{eq:Ideal-LowPass%
X}). By the above, the Fourier transform of this function will s%
imply be $p$ inside the first Brillouin zone, and zero outside i%
t, which is exactly what we want. Now, it is an elementary task 
to compute the derivative of (\protect\ref{eq:Ideal-LowPassX}): 
we find \setcounter{Ceqindent}{0}\protect\begin{eqnarray}d_{\mbox
{\scriptsize$\:\!$ideal}}(x)\hspace{-1.3ex}&\displaystyle\equiv&%
\hspace{-1.3ex}-\mbox{$\protect\displaystyle\protect\frac{d}{dx}%
$}\!\protect\left\{\mbox{$\protect\displaystyle\protect\frac{1}{%
a}$}^{\,}\mbox{sinc}\!\protect\left(\mbox{$\protect\displaystyle
\protect\frac{x}{a}$}\protect\right)\!\protect\right\}=-\mbox{$%
\protect\displaystyle\protect\frac{1}{\pi}$}\mbox{$\protect
\displaystyle\protect\frac{d}{dx}$}\!\protect\left\{\mbox{$%
\protect\displaystyle\protect\frac{\sin(\pi x/a)}{x}$}\protect
\right\}\protect\nonumber\setlength{\Cscr}{\value{CEht}\Ctenthex
}\addtolength{\Cscr}{-1.0ex}\protect\raisebox{0ex}[\value{CEht}%
\Ctenthex][\Cscr]{}\\*[0ex]\protect\displaystyle\hspace{-1.3ex}&%
\displaystyle=&\hspace{-1.3ex}-\mbox{$\protect\displaystyle
\protect\frac{\cos(\pi x/a)}{ax}$}+\mbox{$\protect\displaystyle
\protect\frac{\sin(\pi x/a)}{\pi x^2}$}.\protect\label{eq:Ideal-%
DIdeal}\setlength{\Cscr}{\value{CEht}\Ctenthex}\addtolength{\Cscr
}{-1.0ex}\protect\raisebox{0ex}[\value{CEht}\Ctenthex][\Cscr]{}%
\protect\end{eqnarray}\setcounter{CEht}{10}This function looks a 
little unfamiliar, but it will be of crucial importance for us, 
so it is worth spending a little time understanding it. As with 
the sinc function, its behaviour\ around \mbox{$\protect
\displaystyle x=0$} takes a little work. We first combine the tw%
o terms over a common denominator: \setcounter{Ceqindent}{0}%
\protect\begin{eqnarray}\hspace{-1.3ex}&\displaystyle d_{\mbox{%
\scriptsize$\:\!$ideal}}(x)=\mbox{$\protect\displaystyle\protect
\frac{-\pi x\cos(\pi x/a)+a\sin(\pi x/a)}{a\pi x^2}$}.\protect
\nonumber\setlength{\Cscr}{\value{CEht}\Ctenthex}\addtolength{%
\Cscr}{-1.0ex}\protect\raisebox{0ex}[\value{CEht}\Ctenthex][\Cscr
]{}\protect\end{eqnarray}\setcounter{CEht}{10}We now expand the 
trigonometric functions as Taylor series: \setcounter{Ceqindent}%
{0}\protect\begin{eqnarray}d_{\mbox{\scriptsize$\:\!$ideal}}(x)%
\hspace{-1.3ex}&\displaystyle=&\hspace{-1.3ex}\mbox{$\protect
\displaystyle\protect\frac{1}{a\pi x^2}$}\!\protect\left\{-\pi x%
\!\protect\left[1-\mbox{$\protect\displaystyle\protect\frac{1}{2%
}$}\!\protect\left(\mbox{$\protect\displaystyle\protect\frac{\pi
x}{a}$}\protect\right)^{\!2}\!\!+{\cal O}(x^4)\protect\right]+a%
\!\protect\left[\mbox{$\protect\displaystyle\protect\frac{\pi x}%
{a}$}-\mbox{$\protect\displaystyle\protect\frac{1}{6}$}\!\protect
\left(\mbox{$\protect\displaystyle\protect\frac{\pi x}{a}$}%
\protect\right)^{\!3}\!\!+{\cal O}(x^5)\protect\right]\protect
\right\}\protect\nonumber\setlength{\Cscr}{\value{CEht}\Ctenthex
}\addtolength{\Cscr}{-1.0ex}\protect\raisebox{0ex}[\value{CEht}%
\Ctenthex][\Cscr]{}\\*[0ex]\protect\displaystyle\hspace{-1.3ex}&%
\displaystyle=&\hspace{-1.3ex}\mbox{$\protect\displaystyle
\protect\frac{1}{a\pi x^2}$}\!\protect\left\{-\pi x+\mbox{$%
\protect\displaystyle\protect\frac{\pi^3x^3}{2a^2}$}+{\cal O}(x^%
5)+\pi x-\mbox{$\protect\displaystyle\protect\frac{\pi^3x^3}{6a^%
2}$}+{\cal O}(x^5)\protect\right\}\protect\nonumber\setlength{%
\Cscr}{\value{CEht}\Ctenthex}\addtolength{\Cscr}{-1.0ex}\protect
\raisebox{0ex}[\value{CEht}\Ctenthex][\Cscr]{}\\*[0ex]\protect
\displaystyle\hspace{-1.3ex}&\displaystyle=&\hspace{-1.3ex}\mbox
{$\protect\displaystyle\protect\frac{\pi^2x}{3a^3}$}+{\cal O}(x^%
3).\protect\label{eq:Ideal-DIdealXZero}\setlength{\Cscr}{\value{%
CEht}\Ctenthex}\addtolength{\Cscr}{-1.0ex}\protect\raisebox{0ex}%
[\value{CEht}\Ctenthex][\Cscr]{}\protect\end{eqnarray}\setcounter
{CEht}{10}Thus the function is perfectly well-behaved (and vanis%
hes linearly) around \mbox{$\protect\displaystyle x=0$}, as woul%
d be expected from the properties of the sinc function itself.%
\par According to the philosophy outlined above, it is the low-p%
ass-filtered derivative function, \mbox{$\protect\displaystyle d%
_{\mbox{\scriptsize$\:\!$ideal}}(x)$}, that we should seek to ``%
sample'' onto the lattice. So let us immediately proceed to do s%
o, by considering $x$-values \mbox{$\protect\displaystyle x_{n\!%
}\equiv na$} corresponding to lattice sites labelled by the inte%
ger~$n$. {}From\ (\protect\ref{eq:Ideal-DIdeal}), and multiplyin%
g by the cell width of $a$, we find \setcounter{Ceqindent}{0}%
\protect\begin{eqnarray}\hspace{-1.3ex}&\displaystyle{\protect\it
\Delta\!\:}_{\mbox{\scriptsize ideal}}(x_n)\equiv a^{\:\!}d_{%
\mbox{\scriptsize$\:\!$ideal}}(x_n)=\mbox{$\protect\displaystyle
\protect\frac{-\pi n\cos n\pi+\sin n\pi}{\pi an^2}$}.\protect
\nonumber\setlength{\Cscr}{\value{CEht}\Ctenthex}\addtolength{%
\Cscr}{-1.0ex}\protect\raisebox{0ex}[\value{CEht}\Ctenthex][\Cscr
]{}\protect\end{eqnarray}\setcounter{CEht}{10}{}From\ (\protect
\ref{eq:Ideal-DIdealXZero}) we have already established that 
\mbox{$\protect\displaystyle{\protect\it\Delta\!\:}_{\mbox{%
\scriptsize ideal}}(x_0)=0$}, so we need simply consider the rem%
aining cases \mbox{$\protect\displaystyle n\neq0$}. We now note 
that \mbox{$\protect\displaystyle\sin n\pi=0$} and \mbox{$%
\protect\displaystyle\cos n\pi=(-1)^n$}\mbox{$\!$}, so that we o%
btain the remarkably simply result \setcounter{Ceqindent}{0}%
\protect\begin{eqnarray}\protect\left.\protect\begin{array}{rcl}%
\protect\displaystyle\hspace{-1.3ex}&\protect\displaystyle{%
\protect\it\Delta\!\:}_{\mbox{\scriptsize ideal}}(x_n)=\left\{%
\begin{array}{ll}0&\mbox{if $n=0$,}\\-(-1)^n\!/an&\mbox{otherwis%
e.}\\\end{array}\right.\setlength{\Cscr}{\value{CEht}\Ctenthex}%
\addtolength{\Cscr}{-1.0ex}\protect\raisebox{0ex}[\value{CEht}%
\Ctenthex][\Cscr]{}\protect\end{array}\protect\right.\protect
\label{eq:Ideal-DeltaIdeal}\protect\end{eqnarray}\setcounter{CEh%
t}{10}What does this mean? It means that the ideal way to comput%
e the first derivative on the lattice is not to use the na\-\"\i
ve\ prescription (\protect\ref{eq:Doubling-Naive}), but rather 
\setcounter{Ceqindent}{0}\protect\begin{eqnarray}\protect\left.%
\protect\begin{array}{rcl}\protect\displaystyle\mbox{$\protect
\displaystyle\protect\frac{df}{dx}$}\hspace{-1.3ex}&\protect
\displaystyle\equiv&\hspace{-1.3ex}\protect\displaystyle\lim_{a%
\rightarrow0}\mbox{$\protect\displaystyle\protect\frac{1}{a}$}%
\setcounter{Cbscurr}{25}\setlength{\Cscr}{\value{Cbscurr}%
\Ctenthex}\addtolength{\Cscr}{-1.0ex}\protect\raisebox{0ex}[%
\value{Cbscurr}\Ctenthex][\Cscr]{}\hspace{-0ex}{\protect\left\{%
\setlength{\Cscr}{\value{Cbscurr}\Ctenthex}\addtolength{\Cscr}{-%
1.0ex}\protect\raisebox{0ex}[\value{Cbscurr}\Ctenthex][\Cscr]{}%
\protect\right.}\hspace{-0.25ex}\setlength{\Cscr}{\value{Cbscurr%
}\Ctenthex}\addtolength{\Cscr}{-1.0ex}\protect\raisebox{0ex}[%
\value{Cbscurr}\Ctenthex][\Cscr]{}\setcounter{CbsD}{\value{CbsC}%
}\setcounter{CbsC}{\value{CbsB}}\setcounter{CbsB}{\value{CbsA}}%
\setcounter{CbsA}{\value{Cbscurr}}\ldots-\mbox{$\protect
\displaystyle\protect\frac{1}{4}$}f(x+4a)+\mbox{$\protect
\displaystyle\protect\frac{1}{3}$}f(x+3a)-\mbox{$\protect
\displaystyle\protect\frac{1}{2}$}f(x+2a)\setcounter{Ceqindent}{%
200}\setlength{\Cscr}{\value{CEht}\Ctenthex}\addtolength{\Cscr}{%
-1.0ex}\protect\raisebox{0ex}[\value{CEht}\Ctenthex][\Cscr]{}\\*%
[0.55ex]\protect\displaystyle\hspace{-1.3ex}&\protect
\displaystyle&\hspace{-1.3ex}\protect\displaystyle{\protect\mbox
{}}\hspace{\value{Ceqindent}\Ctenthex}\setcounter{CEht}{30}+f(x+%
a)-f(x-a)\setcounter{Ceqindent}{150}\setlength{\Cscr}{\value{CEh%
t}\Ctenthex}\addtolength{\Cscr}{-1.0ex}\protect\raisebox{0ex}[%
\value{CEht}\Ctenthex][\Cscr]{}\\*[0.55ex]\protect\displaystyle
\hspace{-1.3ex}&\protect\displaystyle&\hspace{-1.3ex}\protect
\displaystyle{\protect\mbox{}}\hspace{\value{Ceqindent}\Ctenthex
}+\mbox{$\protect\displaystyle\protect\frac{1}{2}$}f(x-2a)-\mbox
{$\protect\displaystyle\protect\frac{1}{3}$}f(x-3a)+\mbox{$%
\protect\displaystyle\protect\frac{1}{4}$}f(x-4a)-\ldots
\setlength{\Cscr}{\value{CbsA}\Ctenthex}\addtolength{\Cscr}{-1.0%
ex}\protect\raisebox{0ex}[\value{CbsA}\Ctenthex][\Cscr]{}\hspace
{-0.25ex}{\protect\left.\setlength{\Cscr}{\value{CbsA}\Ctenthex}%
\addtolength{\Cscr}{-1.0ex}\protect\raisebox{0ex}[\value{CbsA}%
\Ctenthex][\Cscr]{}\protect\right\}}\hspace{-0ex}\setlength{\Cscr
}{\value{CbsA}\Ctenthex}\addtolength{\Cscr}{-1.0ex}\protect
\raisebox{0ex}[\value{CbsA}\Ctenthex][\Cscr]{}\setcounter{CbsA}{%
\value{CbsB}}\setcounter{CbsB}{\value{CbsC}}\setcounter{CbsC}{%
\value{CbsD}}\setcounter{CbsD}{1},\setlength{\Cscr}{\value{CEht}%
\Ctenthex}\addtolength{\Cscr}{-1.0ex}\protect\raisebox{0ex}[%
\value{CEht}\Ctenthex][\Cscr]{}\protect\end{array}\protect\right
.\protect\label{eq:Ideal-Ideal}\protect\end{eqnarray}\setcounter
{CEht}{10}which is the SLAC derivative in position space in clos%
ed form. We can recognise the na\-\"\i ve\ derivative operator c%
ontained in the middle two terms of this expression, but with tw%
ice the usual coefficient. The analysis above tells us that it i%
s the omission of all of the other terms that causes the Fourier 
transform of the na\-\"\i ve\ operator to pick up the pathologie%
s of fermion doubling and mirror states.\par As noted above, the 
structure of this operator will be of concern to some. The first 
question that might be asked, however, is simply this: Where on 
Earth did all of those other terms really come from? We started 
with a representation of the Dirac delta function that was nonze%
ro for only one lattice site. Somehow, when we differentiated th%
is function, we ended up with contributions on \mbox{}\protect\/%
{\protect\em all\protect\/} lattice sites (except, perhaps ironi%
cally, the central site itself). How can differentiating ``nothi%
ng'' give us ``something''?\par The answer is, again, contained 
in the careful way that we first passed the formalism through th%
e low-pass filter, to contain it within the first Brillouin zone%
. The resulting sinc function vanished at all of the lattice sit%
es except the central one; but \mbox{}\protect\/{\protect\em its 
derivative is nonzero at all of these other sites\protect\/}. It 
is almost like the sinc function was ``hiding'' the bulk of itse%
lf from our latticised view---but we ``revealed'' it by applying 
the derivative operator, which effectively ``slides'' the functi%
on right and left and reveals its variation to us. This physical 
picture hardly needs elaborating, since the variational formulat%
ion of mechanics (upon which essentially all lattice calculation%
s in quantum physics are based) is rooted in precisely such a co%
nceptual framework!\par Of greater concern to some will be the f%
act that the operation (\protect\ref{eq:Ideal-DeltaIdeal}) is ve%
ry ``nonlocal''\mbox{$\!$}. Of course, different workers in the 
field have concocted conflicting definitions of ``locality'' in 
the context of the lattice---a topic I shall return to in greate%
r detail in the next section---but I believe that by \mbox{}%
\protect\/{\protect\em any\protect\/} of these prior definitions 
of ``locality'' the operation (\protect\ref{eq:Ideal-DeltaIdeal}%
) would be considered ``very nonlocal''\mbox{$\!$}. In absolute 
value, the terms fall off extremely slowly---like $1/x$. The osc%
illatory nature of the signs of the terms means that they might, 
in a sense, be considered to fall off a little more quickly, for 
the same reason that the sum of $1/n$ is logarithmically diverge%
nt but the sum of \mbox{$\protect\displaystyle(-1)^n\!/n$} is co%
nvergent. However, to compute the derivative of some other funct%
ion $f(x)$ at any particular point $x$, we still need to take in%
to account the nature of the function $f(x)$ itself. Clearly, if 
it were to itself, say, diverge linearly for large $x$, the sum 
represented by (\protect\ref{eq:Ideal-DeltaIdeal}) would not be 
convergent (the terms would oscillate between finite values). Bu%
t such divergent functions will not (usually) be within the doma%
in of functions that we wish to consider (and if they are, we wi%
ll in any case find that the definition of the Fourier transform 
will probably break down, or at least require medical attention)%
.\par The more serious questions are the following: Would a ``no%
nlocal'' operator such as (\protect\ref{eq:Ideal-DeltaIdeal}) de%
stroy the structure of manifestly local field theories? And how 
could anyone possibly implement the operator (\protect\ref{eq:Id%
eal-DeltaIdeal}) without an explosion in computational cost? The%
se two questions will be dealt with in the next two sections, re%
spectively.\par\refstepcounter{section}\vspace{1.5\baselineskip}%
\par{\centering\bf\thesection. ``Local'' and ``nonlocal'' can be 
misnomers on the lattice\\*[0.5\baselineskip]}\protect\indent
\label{sect:Local}Given the fundamental importance of locality i%
n (continuum) interacting field theory, it is understandable tha%
t lattice workers are concerned that this property is not violat%
ed on the lattice. However, this concern has led, in my opinion, 
to a number of misconceptions, which I believe need to be correc%
ted.\par The na\-\"\i ve\ derivative operator (\protect\ref{eq:D%
oubling-Naive}) is generally described as being ``local''\mbox{$%
\!$}. The main justification for this description is that it onl%
y involves neighbouring lattice sites to the site in question, a%
nd so as the lattice spacing is taken to zero in the continuum l%
imit, the points essentially become coincident.\par A common gen%
eralisation of this definition of ``locality'' is to deem an ope%
rator ``local'' if it involves only a \mbox{}\protect\/{\protect
\em finite number\protect\/} of neighbouring lattice sites. Agai%
n, the idea is that a finite number of lattice spacings, multipl%
ied by an infinitesimally shrinking lattice spacing distance, eq%
uates to an infinitesimal distance, and hence a ``local'' operat%
ion.\par There are several problems with such definitions.\par F%
irstly, it is generally assumed that any operator \mbox{}\protect
\/{\protect\em not\protect\/} satisfying any particular definiti%
on in question is, in fact, ``nonlocal'' in the continuum limit. 
This conclusion is too harsh. What one must do is determine whet%
her, as the lattice spacing is shrunk in real space, the definit%
ion in question ``picks out'' the derivative of any suitably dif%
ferentiable test function at the point in question. Clearly, ope%
rators only employing a finite number of neighbours will satisfy 
this requirement. \mbox{}\protect\/{\protect\em But the converse 
is not true:\protect\/} an infinite number of lattice sites may 
be involved in the operation, and the operation may still be loc%
al in the continuum limit, provided that, as the lattice spacing 
is reduced, the contribution from any \mbox{}\protect\/{\protect
\em finite interval\protect\/} at any \mbox{}\protect\/{\protect
\em finite distance\protect\/} from the point in question vanish%
es in the limit. It can be shown that the ideal SLAC operator de%
scribed in \mbox{Sec.~$\:\!\!$}\protect\ref{sect:Ideal} satisfie%
s this requirement (for suitably non-divergent test functions, a%
t any rate).\par The second fundamental problem with conventiona%
l definitions of a ``local'' operator on the lattice is that the 
impression is sometimes conveyed that the operator in question 
\mbox{}\protect\/{\protect\em is\protect\/} actually local \mbox
{}\protect\/{\protect\em for a finite lattice spacing distance%
\protect\/}. This cannot, of course, be true. For a finite latti%
ce spacing, we are undeniably linking fields at one lattice poin%
t to fields at other lattice points, which are finite distances 
away. These are manifestly nonlocal interactions. The point, of 
course, is that we wish to retrieve the locality of interactions 
\mbox{}\protect\/{\protect\em in the continuum limit\protect\/}. 
Locality is one property of the continuum formalism that we cann%
ot, by any trick, observe for finite lattice spacing.\par This i%
s made clearer if we consider a Taylor series expansion of the t%
est function $f(x)$ about any point $x$ in the continuum. Clearl%
y, to find \mbox{$\protect\displaystyle f(x\pm a)$} at some fini%
te distance $a$ away from $x$, we would need to know the values 
of \mbox{}\protect\/{\protect\em all\protect\/} of the (infinite 
number of) derivatives of $f$ at the point~$x$. By a reversion o%
f series, the converse is equally true: to find the value of 
\mbox{$\protect\displaystyle f^{\prime\:\!\!}(x)$} using only th%
e values of $f$ on lattice sites $x_n$, it is necessary to use 
\mbox{}\protect\/{\protect\em all\protect\/} of the (infinite nu%
mber of) values of $f$ (for an infinite lattice). In effect, we 
can use the na\-\"\i ve\ operator (\protect\ref{eq:Doubling-Naiv%
e}) to give us a first-approximation to the derivative, but then 
we would need to use the higher-order finite differences in orde%
r to correct the contribution of the higher derivatives to the T%
aylor series expansion.\par This is precisely what is being achi%
eved by the ideal SLAC operator (\protect\ref{eq:Ideal-Ideal}).%
\par\refstepcounter{section}\vspace{1.5\baselineskip}\par{%
\centering\bf\thesection. The proposed derivative operator\\*[0.%
5\baselineskip]}\protect\indent\label{sect:Proposed}The final---%
and valid---objection to the SLAC lattice derivative operator (%
\protect\ref{eq:Ideal-Ideal}) is a practical one. How can one po%
ssibly implement it without blowing out the computational cost e%
normously? The operator (\protect\ref{eq:Ideal-Ideal}) links a g%
iven (one-dimensional) lattice site to every other site in the l%
attice. Compare this to the na\-\"\i ve\ operator (\protect\ref{%
eq:Doubling-Naive}), which only links the given site to its two 
nearest neighbours. The time required to perform such a computat%
ion will increase by a factor of half the number of lattice site%
s!\par The solution I propose to this problem recognises the nat%
ure of the calculations that we wish to perform on the lattice. 
We are not interested in looking at the numerical result of each 
and every application of the derivative operator to a given fiel%
d. Rather, our calculations are such that we need to perform the 
derivative operation many, many times. These results are thrown 
together with the dynamics of the system we are trying to model, 
and at the end of the day we extract a small set of numbers that 
provide a good estimate (we hope) of some physical property of t%
he system in question.\par In effect, each such extracted value 
represents the integrating up of a huge number of elementary ope%
rations. In this context, it is reasonable to consider allowing 
some statistical uncertainty in the definition of each elementar%
y operation. Even if each individual application of the operatio%
n does not manifestly exhibit the properties of the ideal SLAC o%
perator, we can be confident that, \mbox{}\protect\/{\protect\em
on the average\protect\/}, the ideal operator will be faithfully 
represented. To do this, we need simply ensure that the \mbox{}%
\protect\/{\protect\em expectation values\protect\/} of the prop%
erties of the individual operators being applied are equal to th%
e properties of the desired ideal SLAC operator.\par That is the 
general philosophy. Let me now make it concrete. Look back at th%
e SLAC derivative operator (\protect\ref{eq:Ideal-Ideal}). Insid%
e it is contained twice the na\-\"\i ve\ derivative operator (%
\protect\ref{eq:Doubling-Naive}) that uses lattice sites that ar%
e one lattice spacing away from the point in question. My first 
stipulation is that, \mbox{}\protect\/{\protect\em every\protect
\/} time the derivative operator is applied, we \mbox{}\protect
\/{\protect\em at least\protect\/} compute twice the contributio%
n due to the na\-\"\i ve\ operator. This provides our starting p%
oint. We are thus doing no worse, in some sense, than what we wo%
uld do with the na\-\"\i ve\ operator.\par We next consider the 
terms that involve sites that are \mbox{}\protect\/{\protect\em
two\protect\/} lattice spacings away. Apart from an extra minus 
sign, these two terms come in with a factor of $1/2$. Now, inste%
ad of calculating this difference, and multiplying it by $1/2$, 
I make the following stipulation: by random selection, only calc%
ulate this difference \mbox{}\protect\/{\protect\em half of the 
time\protect\/}. If you calculate it, add it in completely. If y%
ou don't calculate it, don't add anything in.\par We next turn t%
o the two terms involving sites that are \mbox{}\protect\/{%
\protect\em three\protect\/} lattice spacings away. The sign has 
flipped back to positive, and the terms come in with a coefficie%
nt of $1/3$. Again, by random selection (independent of the abov%
e selection!), calculate this difference only \mbox{}\protect\/{%
\protect\em one-third\protect\/} of the time, and add it in (com%
pletely) only if you do calculate it.\par Continue this process, 
until you have gone right up to the limit of sites that are half 
a lattice away (using periodic boundary conditions where necessa%
ry).\par The sum of the terms that you \mbox{}\protect\/{\protect
\em did\protect\/} calculate is now the value that you use for t%
he derivative.\par How much of a factor increase in computationa%
l time does such a proposal entail? In terms of the actual compu%
tation of differences---the costly part---the number of calculat%
ions you will need to perform, on average, will go like the \mbox
{}\protect\/{\protect\em logarithm\protect\/} of the number of l%
attice sites, because it is simply \mbox{$\protect\displaystyle1%
+1/2+1/3+\ldots$} up to half the number of sites, which (like th%
e integral of $1/x$) is logarithmic. This is a lot better than b%
eing proportional to the number of lattice sites, as a na\-\"\i
ve\ application of the SLAC derivative would entail!\par The fac%
tor is not convergent, unfortunately, as the number of lattice s%
ites goes to infinity; to make a convergent derivative operator 
of this sort, one would have to reduce the probabilities of the 
higher-distance differences, in some functional way, but compens%
ate by multiplying each such difference by an increasingly large 
coefficient. Such a scenario sounds inherently unstable; and it 
may be quite possibly to prove that the fluctuations inherent in 
such a process would not reduce in variance over a large ensembl%
e of such operations. In any case, it sounds like a bad way to p%
roceed to me; living with a factor that is only logarithmic in t%
he number of lattice sites is quite reasonable in any practical 
context.\par The only remaining concern is the task of randomisi%
ng the decision about whether to compute each difference at a gi%
ven distance. Clearly, this selection task is the only part of t%
he calculation that is \mbox{}\protect\/{\protect\em not\protect
\/} logarithmically sped up, because for each distance, we need 
to decide whether to compute or not compute. It would be nice if 
there were some simple way to spit out random numbers that told 
us \mbox{}\protect\/{\protect\em which\protect\/} distances to c%
ompute; but I have not been able to think up any method that is 
quicker than a brute force loop.\par Fortunately, such an explic%
it loop is relatively cheap in computational terms. Imagine we h%
ave at our disposal a (pseudo-)random bitstream. To decide wheth%
er to perform the distance-2 difference, we simply roll one bit 
off this stream. If it is 0, we do it; if it is 1, we don't. For 
distance-3, we roll off two bits. If the binary number created b%
y these two bits is greater than 2, we try again. Otherwise, we 
check if the number is 0; if so, we do it; if it is 1 or 2, we d%
on't. Likewise, for distance-4; we never need to re-try. We cont%
inue on in this fashion. Now, generating a pseudorandom bitstrea%
m takes only a simple addition per machine integer; comparing to 
the loop variable is a single instruction in machine code; and t%
esting for zero is another single instruction in machine code. I%
ncrementing the loop counter takes one further instruction. Thus%
, even though, using this approach, we need to loop through ever%
y single lattice site, the number of machine cycles (if written 
in assembly language) needed to make the decisions is four times 
the number of lattice sites in one dimension, divided by some ef%
ficiency factor (better than one-half!)\ for the case of discard%
ed values. In most practical cases this should be a negligibly s%
mall cost compared to the difference operations themselves. (In 
cases where even this cost is prohibitive, it may be possible to 
``precompute'' a large set of yes--no flags in a lookup table th%
at could be re-used.)\par\refstepcounter{section}\vspace{1.5%
\baselineskip}\par{\centering\bf\thesection. Application to the 
second-derivative\\*[0.5\baselineskip]}\protect\indent\label{sec%
t:Second}Although, as noted earlier, it does not suffer from the 
pathologies of the fermion doubling and ``mirror states'' proble%
ms, the \mbox{}\protect\/{\protect\em second\protect\/}-derivati%
ve operator is nevertheless of fundamental interest to gauge the%
ory calculations. If one is already applying the first-derivativ%
e operator of the previous section to one's lattice formalism, o%
ne might ask whether a similar improvement to the action may be 
possible by applying the same principles to the second-derivativ%
e.\par I believe that it can. All that we need to do is take the 
negative of the derivative of the ideal (continuum but low-pass-%
filtered) derivative operation (\protect\ref{eq:Ideal-DIdeal}): 
\setcounter{Ceqindent}{0}\protect\begin{eqnarray}d^{\:\!(2)}_{%
\mbox{\scriptsize$\:\!$ideal}}(x)\hspace{-1.3ex}&\displaystyle
\equiv&\hspace{-1.3ex}-\mbox{$\protect\displaystyle\protect\frac
{d}{dx}$}^{\:\!}d_{\mbox{\scriptsize$\:\!$ideal}}(x)\protect
\nonumber\setlength{\Cscr}{\value{CEht}\Ctenthex}\addtolength{%
\Cscr}{-1.0ex}\protect\raisebox{0ex}[\value{CEht}\Ctenthex][\Cscr
]{}\\*[0ex]\protect\displaystyle\hspace{-1.3ex}&\displaystyle=&%
\hspace{-1.3ex}-\mbox{$\protect\displaystyle\protect\frac{2\cos(%
\pi x/a)}{ax^2}$}+\mbox{$\protect\displaystyle\protect\frac{2\sin
(\pi x/a)}{\pi x^3}$}-\mbox{$\protect\displaystyle\protect\frac{%
\pi\sin(\pi x/a)}{a^2x}$}.\protect\label{eq:Second-D2Ideal}%
\setlength{\Cscr}{\value{CEht}\Ctenthex}\addtolength{\Cscr}{-1.0%
ex}\protect\raisebox{0ex}[\value{CEht}\Ctenthex][\Cscr]{}\protect
\end{eqnarray}\setcounter{CEht}{10}Again, when we ``sample'' ont%
o the lattice, and multiply by the size $a$ of the lattice cell, 
we obtain \setcounter{Ceqindent}{0}\protect\begin{eqnarray}%
\hspace{-1.3ex}&\displaystyle{\protect\it\Delta\!\:}^{(2)}_{\mbox
{\scriptsize ideal}}(x_n)\equiv a^{\:\!}d^{\:\!(2)}_{\mbox{%
\scriptsize$\:\!$ideal}}(x_n)=-\mbox{$\protect\displaystyle
\protect\frac{2\cos n\pi}{a^2n^2}$}+\mbox{$\protect\displaystyle
\protect\frac{2\sin n\pi}{a^2n^3\pi}$}-\mbox{$\protect
\displaystyle\protect\frac{\pi\sin n\pi}{a^2n}$}.\protect
\nonumber\setlength{\Cscr}{\value{CEht}\Ctenthex}\addtolength{%
\Cscr}{-1.0ex}\protect\raisebox{0ex}[\value{CEht}\Ctenthex][\Cscr
]{}\protect\end{eqnarray}\setcounter{CEht}{10}In this case we ne%
ed to be careful when extracting the coefficient at \mbox{$%
\protect\displaystyle n=0$}; in the limit \mbox{$\protect
\displaystyle n\rightarrow0$}, one finds \setcounter{Ceqindent}{%
0}\protect\begin{eqnarray}\hspace{-1.3ex}&\displaystyle{\protect
\it\Delta\!\:}^{(2)}_{\mbox{\scriptsize ideal}}(0)=-\mbox{$%
\protect\displaystyle\protect\frac{\pi^2}{3a^2}$}.\protect
\nonumber\setlength{\Cscr}{\value{CEht}\Ctenthex}\addtolength{%
\Cscr}{-1.0ex}\protect\raisebox{0ex}[\value{CEht}\Ctenthex][\Cscr
]{}\protect\end{eqnarray}\setcounter{CEht}{10}For \mbox{$\protect
\displaystyle n\neq0$}, one can again use the identities \mbox{$%
\protect\displaystyle\sin n\pi=0$} and \mbox{$\protect
\displaystyle\cos n\pi=(-1)^n$}\mbox{$\!$}. The overall result f%
or the closed form of the ``second SLAC derivative'' is then 
\setcounter{Ceqindent}{0}\protect\begin{eqnarray}\protect\left.%
\protect\begin{array}{rcl}\protect\displaystyle\hspace{-1.3ex}&%
\protect\displaystyle{\protect\it\Delta\!\:}^{(2)}_{\mbox{%
\scriptsize ideal}}(x_n)=\left\{\begin{array}{ll}-\pi^2\!/3a^2&%
\mbox{if $n=0$,}\\-2(-1)^n\!/a^2n^2&\mbox{otherwise.}\\\end{arra%
y}\right.\setlength{\Cscr}{\value{CEht}\Ctenthex}\addtolength{%
\Cscr}{-1.0ex}\protect\raisebox{0ex}[\value{CEht}\Ctenthex][\Cscr
]{}\protect\end{array}\protect\right.\protect\label{eq:Second-De%
lta2Ideal}\protect\end{eqnarray}\setcounter{CEht}{10}In this cas%
e, the fall-off of the coefficients goes like $1/n^2$ rather tha%
n $1/n$. The operator (\protect\ref{eq:Second-Delta2Ideal}) does 
not contain precisely the na\-\"\i ve\ operator (\protect\ref{eq%
:Doubling-SecondDeriv}), but the structure of the three middle t%
erms is similar, even if the numerical coefficients differ.\par
To implement the operator (\protect\ref{eq:Second-Delta2Ideal}) 
in the ``stochastic'' fashion proposed in \mbox{Sec.~$\:\!\!$}%
\protect\ref{sect:Proposed}, it would probably be sufficient to 
calculate the three middle terms for every application of the de%
rivative, and then to apply the higher-distance contributions (s%
ums, now, rather than differences, but with the correct sign) wi%
th a statistical weight equal to $2/n^2$.\par\refstepcounter{sec%
tion}\vspace{1.5\baselineskip}\par{\centering\bf\thesection. Con%
clusions\\*[0.5\baselineskip]}\protect\indent\label{sect:Conclus%
ions}In this paper I have tried to explain, in a pedagogical way%
, why the SLAC prescription for the definition of the derivative 
operator is the ``cleanest'' from a fundamental point of view. I 
have shown that by very simple arguments it is possible to obtai%
n the position-space form of the SLAC derivative in closed form. 
I have proposed that, by implementing this operator in a stochas%
tic fashion, it will be possible to gain the advantages of the S%
LAC derivative without significantly more computational cost tha%
n using either the na\-\"\i ve\ derivative or the Wilson prescri%
ption for removing the fermion doubling problem.\par\vspace{1.5%
\baselineskip}\par{\centering\bf Acknowledgments\\*[0.5%
\baselineskip]}\protect\indent I gratefully thank Tien D.~Kieu f%
or introducing me to this problem, the participants at the 2002 
Congress of the Australian Institute of Physics for rekindling m%
y interest in it, and Philippe de~Forcrand and Christian Hoelbli%
ng for providing vital comments on the first version of this pap%
er.\par\end{document}